\documentclass[twocolumn]{aastex63}
\usepackage{amsmath} 
\usepackage{amssymb}
\usepackage[utf8x]{inputenc}

\received{}
\revised{}
\accepted{}
\submitjournal{AJ}

\shorttitle{ Spectroscopic and photometric monitoring of an OH/IR star}
\shortauthors{S. Ghosh et al.}

\begin{document}

\title{ \textbf{\Large Spectroscopic and photometric monitoring of a poorly known high-luminous OH/IR star: IRAS 18278+0931}}

\correspondingauthor{Supriyo Ghosh, Soumen Mondal}
\email{supriyoani89@gmail.com, soumen.mondal@bose.res.in}

\author[0000-0002-0786-7307]{Supriyo Ghosh}
\affil{S. N. Bose National Centre for Basic Sciences, Salt Lake, Kolkata-700 106, India}
\affil{Tata Institute of Fundamental Research, Homi Bhabha Road, Colaba, Mumbai 400 005, India}

\author{Soumen Mondal}
\affil{S. N. Bose National Centre for Basic Sciences, Salt Lake, Kolkata-700 106, India}

\author{Ramkrishna Das}
\affiliation{S. N. Bose National Centre for Basic Sciences, Salt Lake, Kolkata-700 106, India}

\author{Somnath Dutta}
\affiliation{S. N. Bose National Centre for Basic Sciences, Salt Lake, Kolkata-700 106, India}
\affiliation{Academia Sinica Institute of Astronomy and Astrophysics, PO Box 23-141, Taipei 106, Taiwan}


\begin{abstract}
We present the time-dependent properties of a poorly known OH/IR star $-$ IRAS 18278+0931 (hereafter, IRAS 18+09) towards the Ophiuchus constellation. We have carried out long-term optical/near-infrared (NIR) photometric and spectroscopic observations to study the object. From optical $R$- and $I$-band light curves, the period of IRAS 18+09 is estimated to be 575 $\pm$ 30 days and the variability amplitudes ranges from $\Delta$R $\sim$ 4.0 mag to $\Delta$I $\sim$ 3.5 mag. From the standard Period-Luminosity (PL) relations, the distance ($D$) to the object, 4.0$\pm$1.3 kpc, is estimated. Applying this distance in the radiative transfer model, the spectral energy distribution (SED) are constructed from multi-wavelength photometric and IRAS-LRS spectral data which provides the luminosity, optical depth, and gas mass-loss rate (MLR) of the object to be 9600 $\pm$ 500 $L_{\odot}$, 9.1 $\pm$ 0.6 at 0.55 $\mu$m and 1.0$\times$10$^{-6}$ M$_\odot$ yr$^{-1}$, respectively. The current mass of the object infers in the range 1.0$-$1.5$M_\odot$ assuming solar metallicity. Notably, the temporal variation of atomic and molecular features (e.g., TiO, Na I, Ca I, CO, H$_2$O) over the pulsation cycle of the OH/IR star illustrates the sensitivity of the spectral features to the dynamical atmosphere as observed in pulsating AGB stars. 
\end{abstract}

\keywords{infrared:stars -- method: observational -- stars: AGB and post-AGB -- stars:late-type -- stars: variable:general -- techniques: spectroscopic}


\section{Introduction}
 
 OH/IR stars are a class of long-period (several hundred days) large amplitudes (1 mag bolometric) variables with huge MLR ($>$ 10$^{-5}$ M$_\odot$yr$^{-1}$, \citealt{1983aap..127..73}) representing the thermally pulsing asymptotic giant branch (TP-AGB) phase. Due to the dust formation and copious mass-loss (e.g., \citealt{2018aapr..26..1}) of these stars, circumstellar envelops (CSE) develop, which eventually become opaque to visible light \citep{1996aapr...7...97} and emit at infrared (IR) wavelengths \citep{1987aap..186..136}. OH/IR stars are oxygen-rich stars (C/O$<$1, O-rich) and share many common characteristics with O-rich Mira variables, implying a close relationship between them. However, it is not well-understood how they differ from each other (see \citealt{1998aap..329..991} for details). Moreover, O-rich stars at TP-AGB phase evolve from solar metallicity star with either progenitor mass below 2 M$_\odot$ or above 4 M$_\odot$ \citep{2013MNRAS.434..488M}, and they often show OH (at 1612, 1665, and 1667 MHz), H$_2$O (mostly at 22 GHz), and SiO (mostly at 43 and 86 GHz) maser \citep{2012msa..book.....G}. Thus, the OH counterpart of the name comes because of the detectable OH maser emission in those stars \citep{2018mnras..479..3545}, where the thermal radiation emitted by warm dust acts as a pump. The luminosity distribution of OH/IR stars peaks at around 5000$L_\odot$ indicating low initial mass ($<$ 2M$_\odot$) progenitors \citep{1988aap...200...40, 2018mnras..479..3545}. However, an appreciable number of OH/IR stars with a luminosity well-above 10 000$L_\odot$ exist in both Galactic disk (e.g., \citealt{1988aap...200...40}) and bulge (e.g., \citealt{2007MNRAS.381.1219O}), classified as a ``high-luminosity group'' \citep{2015A&A...579A..76J}, representing an evolution from relatively massive progenitors (4--6 M$_\odot$).
 
Most OH/IR stars are detected through either 1612 MHz OH surveys in the Galactic plane and Galactic center (e.g., \citealt{2001A&A...366..481S}) or through various systematic surveys (for example, the Arecibo surveys $-$ \citealt{1988ApJS...66..183E,1990ApJ...362..634L, 1993ApJS...89..189C}), which have been conducted for sources with IRAS (IRAS Point Source Catalogue (1988)) colors resembling those of OH/IR stars (see e.g., \citealt{1984ApJ...278L..41O}). Using the same IRAS color criterion, a sub-class of OH/IR stars were proposed to be pre-planetary nebulae (i.e., in the post-AGB evolutionary phase), as confirmed by an HST imaging survey \citep{2007AJ....134.2200S}. Such studies also resulted in the serendipitous discoveries of luminous objects that were not evolved stars (e.g., \citealt{2013MNRAS.428.1537P}). These previous works demonstrate the importance of studying individual IRAS sources for their characterization. 

The pulsating stellar interiors of AGB stars trigger outward shock fronts, which makes their atmospheres cool and very extended. As a consequence, several observational effects, e.g., modulation of the atmospheric structure, alteration of the chemical composition of the gas, effects of the velocity field in the line formation regions, and the large variation in luminosity, are seen. The shapes of the spectral energy distributions (SEDs) and the depth of the silicate features (at 9.7 and 18 $\mu$m) show significant variation with the pulsation phase (e.g., \citealt{2002A&A...391..665S}). The photometric light curves reflect the variation in brightness, surface temperature, and radius \citep{LeBertre1992}. Thus, a long-term monitoring program is required for studying the variability properties (for e.g., amplitudes and color variation) of this class of stars. In addition, several molecules (e.g., TiO, VO, H$_2$O, and CO) form efficiently in their cool extended atmosphere, resulting in the typical line-rich late-type spectra. While TiO and VO bands are the dominant molecules at the optical spectra, CO and H$_2$O molecules shape the NIR spectrum. Previous studies unveil not only the considerable and independent variation of individual spectral features (optical TiO and VO bands: \citealt{1998aap..330..1109} and NIR CO, OH and H$_2$O lines: \citealt{1982apj..252..697, 1984ApJS...56....1H} or \citealt{1999aap...341..224}) over the phase but also their cycle-to-cycle variation. The line profiles of molecular lines are also affected because of the velocity field variation, while the molecular abundances in the atmosphere are influenced by the luminosity variation \citep{2017A&A...606A...6L, 2018aj..155..216}. Moreover, the NIR spectrum of OH/IR stars is strongly influenced by the dust, OH radical and water content in the outer atmosphere \citep{1996aap..307..481, 2006A&A...455..645V}. Theoretical models of O-rich stars have been successfully applied for studying line profiles variations and radial velocity curves found in observed spectra of long-period variables \citep{2010aap..514..35,2017A&A...606A...6L}, however, no model still can interpret the exact dynamical variation of AGB atmosphere or spectral variations at late spectral type with pulsation. Moreover, time-series spectral data are sparse in the literature, even rarer for an OH/IR star, and much needed to model the complex convective atmosphere of the AGB stars.

\begin{figure}
	\center
	\includegraphics[scale=1.6]{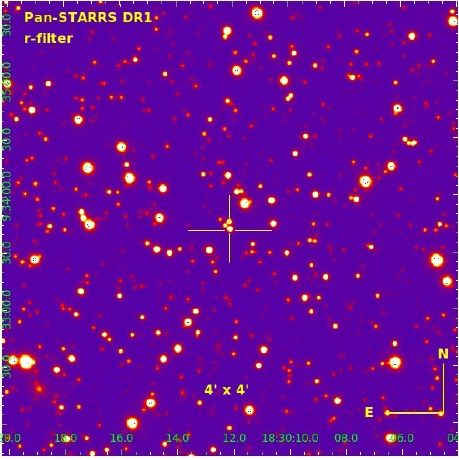}
	\caption{Finder Chart of our source IRAS 18278+0931, coordinates: $\alpha_{(2000)}$ = 18$^h$30$^m$12.10$^s$, $\delta_{(2000)}$ = +09\degr33\arcmin42.6\arcsec.}
	\label{fig1:fc}
\end{figure}
This paper reports spectro-photometric time-series observations of an OH/IR star IRAS 18+09 ($\alpha_{(2000)}$ = 18$^h$30$^m$12.10$^s_{(0.21 mas)}$, $\delta_{(2000)}$ = +09\degr33\arcmin42.6\arcsec$_{(0.23 mas)}$). The finding chart of the object is shown in Figure~\ref{fig1:fc}. The source was detected by 1612 MHz OH maser with Arecibo radio telescope \citep{1984ApJ...278L..41O}. The location of the object in IRAS two-colour ([12]$-$[25] vs. [25]$-$[60]) diagram signifies a more evolved O-rich circumstellar shells \citep{1988A&A...194..125V}. The IRAS LRS spectrum \citep{1986A&AS...65..607O} of the object shows strong silicate dust (O-rich dust) emission feature at 9.7 $\mu$m indicating the population of group E in the LRS spectral classification (see \citealt{1997ApJS..112..557K}). In addition, simultaneous observations of H$_2$O and SiO masers are found recently \citep{2017ApJS..232...13C}. In this paper, we study the variability properties of the source (periods, amplitudes, and color variations) and shed light on the different time dependencies of various spectral signatures as a result of the modulation of the stellar atmosphere caused by the varying luminosity from low-resolution optical/NIR spectra. We also present a comparative study of the time-dependent spectral variations of the OH/IR star with the classical Mira variables. This paper is organized as follows. The observations and data reduction procedures are presented in Section~\ref{observations_data_reduc}. Section~\ref{result_and_discussion} deals with our new results and discussion. Summary and conclusion of our studies are presented in Section~\ref{summary_and_conclusion}.

\section{Observations and Data Reduction} \label{observations_data_reduc}

The optical photometric and spectroscopic observations were carried out using Hanle Faint Object Spectrograph and Camera (HFOSC) on the 2-m Himalayan Chandra Telescope (HCT) at Hanle, India. The source was monitored in the optical $R$- and $I$-band, over 11 epochs during 2014 August 19 $-$ 2018 April 17, using 2K $\times$ 2K HFOSC imaging CCD (field of view of about 10 $\times$ 10 arcmin$^2$ with a plate scale of 0.296 arcsec pixel$^{-1}$). For spectroscopy (7 epochs), we have used Grism no. 8 (Gr\#8, 580$-$900~$nm$) of the instrument with a resolving power of $\approx$ 2200\footnote{\url{https://www.iiap.res.in/iao_hfosc}}.

The NIR photometric observations (1 epoch) were acquired using Near-Infrared Imaging Camera cum Multi-Object Spectrograph (NICMOS-3) on the 1.2-m Mt. Abu telescope, India, and the spectroscopic observations (6 epochs) were done using NICMOS-3 and TIFR Near-Infrared Spectrometer and Imager (TIRSPEC, \citealt{2014JAI.....350006}) on the 2-m HCT covering 1.5--2.4 $\mu$m region. The NICMOS-3 has a 256 $\times$ 256 HgCdTe detector array with a resolution, R $\approx$1000; while TIRSPEC has 1024 $\times$1024 Hawaii-1 array which provides a resolution $\approx$1200. Log of observations are mentioned in Table~\ref{tab:log}. Photometric observations on 2013 May 28 in $JHK^\prime$-bands were taken in five dithered positions to generate the NIR sky frame, and multiple frames are taken in each position to get a better signal to noise ratio (S/N). In the spectroscopic observing mode, the spectra were taken at two positions dithered by $\sim$ 10\arcsec along with the slit to subtract the sky. Several such set of frames were observed to improve S/N. The estimated S/N is $\sim$50$-$80 in $H$-band and  $\sim$70$-$120 in $K$-band for TIRSPEC data, while the S/N is $\sim$30 in $H$ and $K$ for NICMOS-3 data.

The observed data were reduced with the help of standard tasks of the Image Reduction and Analysis Facility (IRAF\footnote{\url{http://iraf.noao.edu/}}) following the standard reduction method. The aperture photometry was performed using the APPHOT package of IRAF. The zero-points of photometry were determined using the standard stars. The time-series $R$ and $I$ aperture magnitudes are listed in Table~\ref{tab:phot}. The errors mentioned in Table~\ref{tab:phot} are purely photometric errors of the object. The systematic errors coming from the standard stars are ignored here. The NICMOS-3 spectroscopic data were analysed using the APALL task of IRAF. The TIRSPEC data were reduced with TIRSPEC pipe-line\footnote{\url{https://github.com/indiajoe/TIRSPEC/wiki}} \citep{2014JAI.....350006} and was cross-checked with the IRAF reduction method. Both methods of reduction match well. The additional details of the reduction process can be found in \citet{2018aj..155..216}.

\begin{table*}
	\begin{center}
		\caption{ Log of photometric and spectroscopic observations}
		\label{tab:log}
		\resizebox{1.0\textwidth}{!}{
			\begin{tabular}{crrrrrrr}
				\hline
				Date of Observation & Observation Type & Spectral Band & Int. Time (s) & No of Frames & Telescope & Remarks\\
				\hline
				\hline
				2013-May-28 & Photometry & $J/H/K$ & 1/1/0.4 &5* [15/15/25]  & 1.2 m Mt. Abu& clear sky\\ 
				\hline
				2014-Aug-19& Photometry& $R/I$ & 80/4 & 3/3 & 2 m HCT & clear sky\\   
				2015-July-05& Photometry & $R/I$ & 150/20 & 3/3 & 2 m HCT & clear sky \\ 
				2015-Oct-06& Photometry & $R/I$ & 150/20 & 2/2 &  2 m HCT & clear sky \\ 
				2015-Nov-19& Photometry & $R/I$ & 150/20 & 1/2 & 2 m HCT & clear sky  \\ 
				2016-Feb-15& Photometry & $R/I$ & 100/12 & 3/3 & 2 m HCT & clear sky \\ 
				2016-May-08& Photometry & $R/I$ & 100/5 & 3/3 & 2 m HCT & clear sky\\ 
				2016-Jun-26& Photometry & $R/I$ & 100/5 & 3/3 & 2 m HCT & clear sky\\   
				2016-Nov-02& Photometry & $R/I$ & 120/7 & 3/3 & 2 m HCT & clear sky\\   
				2017-Apr-07& Photometry & $R/I$ & 60/600 & 2/2 & 2 m HCT & clear sky\\
				2017-Jun-25& Photometry & $R/I$ & 20/500 & 2/2 & 2 m HCT & clear sky\\ 
				2018-Apr-17& Photometry & $R/I$ & 20/400 & 2/2 & 2 m HCT & clear sky\\ 
				\hline
				2013-May-28& Spectroscopy &$H/K/KA$      & 120/120/120 & 2*1 & 1.2 m Mt. Abu & clear sky \\
				2013-Oct-15& Spectroscopy & 600-900~$nm$ & 2400 &1 & 2 m HCT & clear sky\\ 
				2013-Nov-07& Spectroscopy & 600-900~$nm$ & 1800 &1 & 2 m HCT & clear sky\\ 
				2014-Aug-19& Spectroscopy & 600-900~$nm$ & 300 &1 & 2 m HCT & clear sky\\
				2014-Oct-29 & Spectroscopy & $HK$    &100     &2*5   &2 m HCT & clear sky \\
				2015-Mar-02 & Spectroscopy & $HK$    &100     &2*3   &2 m HCT & clear sky \\
				&              & 600-900~$nm$  & 1800 &1      &  &  \\  
				2015-July-05& Spectroscopy & 600-900~$nm$ & 1500 &1 & 2 m HCT & clear sky\\
				2015-Oct-07 & Spectroscopy & $HK$    &100     &2*5   &2 m HCT & clear sky \\
				&              & 600-900~$nm$ & 2000 &1 &  &  \\ 
				2015-Nov-20 & Spectroscopy & $HK$    &100     &2*7   &2 m HCT & clear sky \\
				2016-Feb-15& Spectroscopy & 600-900~$nm$ & 2000 &1 & 2 m HCT & clear sky\\ 
				2017-Apr-07 & Spectroscopy & $HK$    &100     &2*11   &2 m HCT & clear sky \\
				\hline
		\end{tabular}}
	\end{center}
\end{table*}

\begin{table*}
\begin{center}
\caption{Optical $RI$ Photometry}
\label{tab:phot}
\resizebox{1.0\textwidth}{!}{
\begin{tabular}{ccccccc}
\hline
Date of Obs. & Epoch & Optical & Telescope/Instrument & $R$ & $I$  & ($R-I$)   \\
(UT)         & (JD)           & Phase   &                      &(mag)& (mag)& (mag) \\
\hline
\hline
2014 Aug 19.75  & 2456889 & 0.97  & HCT/HFOSC & 13.185 $\pm$ 0.003  & 10.080 $\pm$ 0.002 & 3.105  \\
2015 July 05.67 & 2457209 & 1.44 & HCT/HFOSC & 17.416 $\pm$ 0.018 & 13.243 $\pm$ 0.003 & 4.173  \\
2015 Oct 06.69  & 2457302 & 1.61  & HCT/HFOSC & 16.041 $\pm$ 0.005  & 13.254 $\pm$ 0.004 & 2.787 \\
2015 Nov 19.55  & 2457346 & 1.70  & HCT/HFOSC & 15.337 $\pm$ 0.007 & 12.336 $\pm$ 0.004 & 3.001 \\
2016 Feb 15.98  & 2457434 & 1.90  & HCT/HFOSC & 13.346 $\pm$ 0.005 & 10.353 $\pm$ 0.003 & 2.993 \\
2016 May 08.83  & 2457517 & 2.05  & HCT/HFOSC & 13.469 $\pm$ 0.009 & 10.369 $\pm$ 0.005 & 3.100 \\
2016 Jun 26.83  & 2457566 & 2.12  & HCT/HFOSC & 13.854 $\pm$ 0.003 & 10.682 $\pm$ 0.003 & 3.172 \\
2016 Nov 02.53  & 2457695 & 2.31  & HCT/HFOSC & 15.903 $\pm$ 0.011 & 12.164 $\pm$ 0.003 & 3.739 \\
2017 Apr 07.92  & 2457851 & 2.55  & HCT/HFOSC & 16.580 $\pm$ 0.007 & 13.350 $\pm$ 0.004 & 3.234 \\
2017 Jun 25.65  & 2457930 & 2.72  & HCT/HFOSC & 14.936 $\pm$ 0.005 & 11.975 $\pm$ 0.003 & 2.961 \\
2018 Apr 17.60  & 2458226 & 3.25  & HCT/HFOSC & 14.939 $\pm$ 0.005 & 11.847 $\pm$ 0.004 & 3.092 \\
\hline 
Average & ... &... & ... & 15.0 & 11.8 & 3.2 \\
\hline
\end{tabular}}
\end{center}
\end{table*}

\section{Result and discussion} \label{result_and_discussion}

\subsection{Optical Light Curves and Period} \label{op_lc_and_Period}

\begin{figure}
	\center
    \includegraphics[scale=0.30]{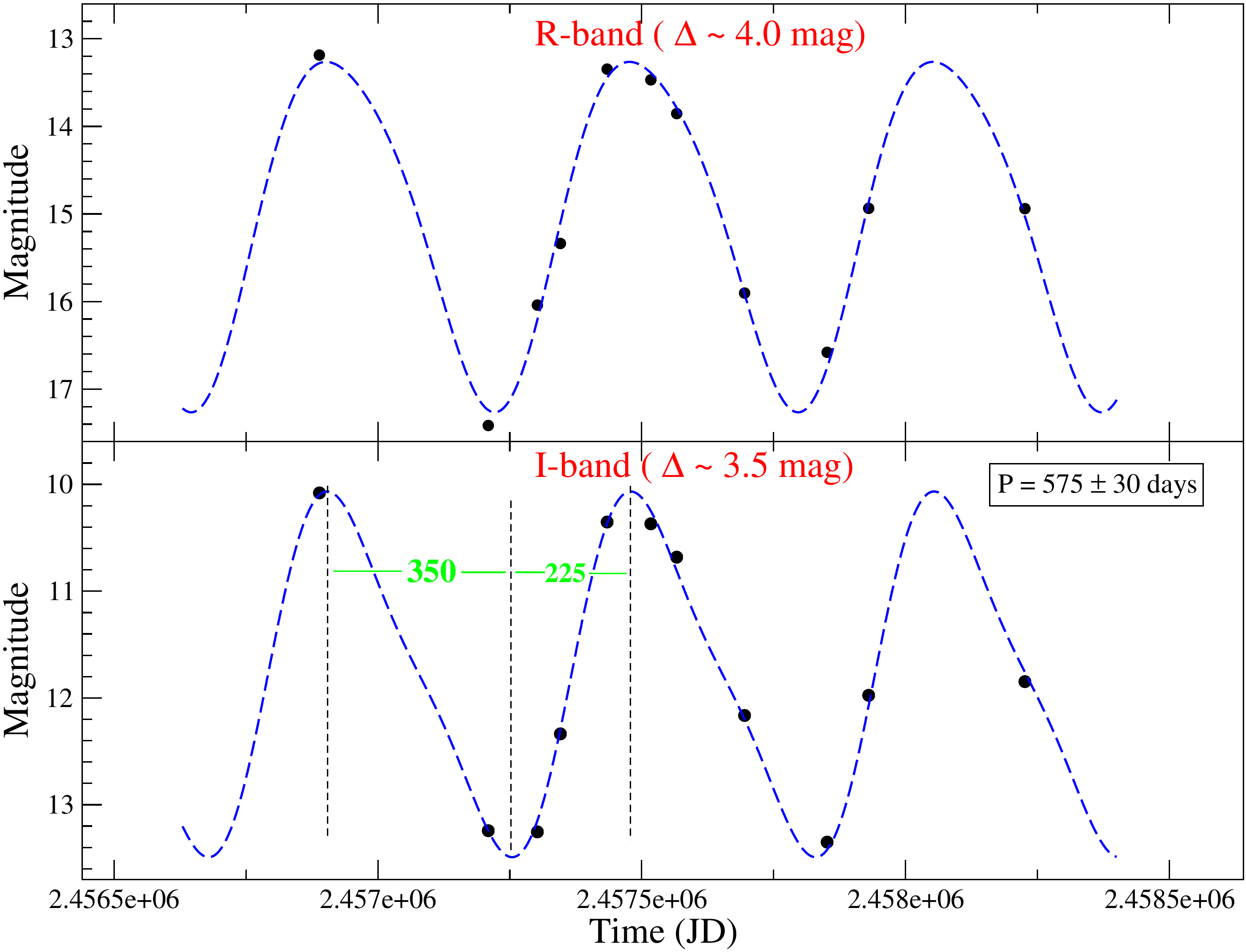}
    \caption{Figure shows the optical light curves of IRAS 18278+0931 in $R$-band (top) and $I$-band (bottom), where the filled circles are our observed data points, while the dashed lines are fitted light curves with P= 575 days. The vertical lines are marked on the maximum and minimum light positions to show the different rise and fall times.}
    \label{fig1:lc}
\end{figure}

 The optical light curves in the $R$- and $I$-band are shown in Figure~\ref{fig1:lc} and time-series $RI$ magnitudes are listed in Table~\ref{tab:phot}. To estimate the period of the object, we applied the Fourier decomposition technique \citep{2013arXiv1309.4297, 2018aj..155..216},
 
 \begin{equation} \label{Equation1:light_curve_fitting}
 m~(t) = A_0 + \sum_{\kappa=1}^{N}  A_\kappa sin(\kappa \omega t + \phi _\kappa)
 \end{equation} 
 
 where $\omega = 2\pi /P$, $P$ is the period in days, $A_\kappa$ and $\phi _\kappa$ represent the amplitude and phase$-$shift for $\kappa ^{th}$-order, respectively, and $N$ is the order of fit. We fit the light curves considering up to second-order terms of the equation, and find the best fit from the $\chi^2$ minimization technique. The best-fit light curve provides a period of 575 $\pm$ 30 days. The amplitude of variability is $\sim$ 4.00 mag in $R$-band and $\sim$ 3.5 mag in $I$-band. We find the greater fall time ($F_t$) of the light curve as 350 days than the rise time ($R_t$) as 225 days, which indicats the asymmetric behavior of the optical light curves. Such behavior was first noticed by \citet{1975A&A....39..473B} in OH emitting Mira variables. Furthermore, it may appear that there is a phase lag between $R$ and $I$-band light curves. However, this result is inconclusive because of a few number of observations. It is to be noted that the provided optical phases in this paper are derived from the asymmetric light curve as described in \citet{1990A&A...239..193V} considering Phase ($\Phi$)= 0 at 2456329 JD.
 
The NIR photometric observations were carried out on 2013 May 28.92 UT ($\Phi$ $\sim$ 0.16) yielding magnitudes of $J$ = 6.46 $\pm$ 0.04 mag, $H$ = 5.15 $\pm$ 0.05 mag, $K$ = 4.37 $\pm$ 0.04 mag. The 2MASS $JHK$ measurements (made on 1999 July 23) yield magnitudes of $J$ = 7.02 $\pm$ 0.03 mag, $H$ = 5.47 $\pm$ 0.03 mag and $K$ = 4.52 $\pm$ 0.02 mag \citep{2003yCat.2246....0}. Significant NIR variability is seen even only in two epoch observations. 

\subsection{Distance and Luminosity} \label{distance_and_luminosity} 

We use the period$-$luminosity (PL) relation of \citet{2010A&A...523A..18D} based on O-rich Galactic Mira variables to derive the distance to the object. Taking the average of $K$-band magnitudes on two epochs as described in Section~\ref{op_lc_and_Period}, the PL relation ($M_K$ = --3.47$_{(\pm (0.19)}$ log $P$ + 0.98$_{(\pm 0.45)}$) yields the absolute $K$-band magnitude of the source to be $M_K$ = $-$8.6 $\pm$ 0.7 mag. For distance estimation, the relation, $m_k - M_K$ = 5log $D$ $-$5 + $A_K$ is used, where $A_K$ is the interstellar extinction. Considering the Galactic interstellar extinction in the direction towards the source, $A_K$ = 0.06 \citep{2011ApJ...737..103}, we obtain the distance, D, to the source as 4.0 $\pm$ 1.3 kpc. The uncertainties in the distance measurement come from the uncertainty of the estimated period, PL relation, photometric error of calculating $K$-band magnitude. However, the strong variability in the $K$-band and the circumstellar extinction of heavily obscured sources are the main constraints to this method of determining the distance of OH/IR star \citep{2010A&A...523A..18D}. We have found that the derived distance based on the PL relation is consistent with the distance ($\sim$ 5.0 kpc) estimated by \citet{2019A&A...628A..94A} using the Gaia DR2 parallax measurement \citep{2018A&A...616A...1G} and photometric catalogs.

We have also estimated the bolometric magnitude $M_{bol}$ using the PL relation ($M_{bol}$ = --1.85$_{(\pm 0.24)}$ -- 2.55$_{(\pm 0.10)}$ log $P$) of \citet{1991mnras..248..276}. The relation provides the bolometric magnitude of $M_{bol}$ = $-$5.19 $\pm$ 0.38 corresponding to the luminosity $\sim$9500L$_\odot$ (taking the solar bolometric magnitude $M_{bol, \odot}$ = 4.75 mag from \citealt{2010A&A...523A..18D}).

\subsection{Spectral Energy Distribution} \label{section:SED}

To fit the SED, we have adopted the radiative transfer code, More of Dusty\footnote{\url{http://homepage.oma.be/marting/codes.html}} (MoD, \citealt{2012aap...543..36}), which considers a slightly updated and modified version of the ``DUSTY" code version, 2.01 \citep{1999astro.ph.10475}, as a subroutine within a minimisation code. The MoD works to find best-fit parameters, e.g., luminosity ($L$), dust optical depth at 0.55 $\mu$m ($\tau_{0.55}$), dust temperature at the inner radius ($T_c$) and slope, $p$, of the density distribution ($\rho$ $\sim$ $r^{-p}$) by using photometric-, spectroscopic- and visibility-data as well as one dimensional intensity profiles in the minimization process. Here, we consider photometric and spectroscopic data as the data of the latter two are unavailable in the literature.

For photometric and spectroscopic data, we made use of broadband optical $I$-band to far-infrared (FIR) multi-wavelength data and IRAS LRS spectra. The fluxes, used here for the SED fit, are obtained from the archival Two Micron All Sky Survey (2MASS; \citealt{2003yCat.2246....0}) at 1.23, 1.66 and 2.16 $\mu$m; the AllWISE Data Release \citep{2013yCat.2328....0} at 3.35, 4.6, 11.6, and 22.1 $\mu$m; the AKARI/IRC all-sky Survey (ISAS/JAXA, 2010) \citep{2010aap...514A...1, 2010yCat.2298....0Y} at 8.61, 18.4 and 90 $\mu$m; and IRAS catalog of Point Sources (IPAC 1986) at 12, 25 and 60 $\mu$m as listed in Table~\ref{tab:archival_phot}. The coordinates in the AKARI/FIS Bright Source Catalog are 5.5\arcsec away from the source coordinates as listed in Table~\ref{tab:archival_phot}. However, the position accuracy of 6\arcsec is recommended for all sources in the catalog for a practical and safe value \citep{2010yCat.2298....0Y}. We also included our three epochs (at maximum, minimum, and middle of the light curve) $I$-band variability data and one epoch $JHK^\prime$ measurements in our calculation. The observed $I$ and $JHK^\prime$ magnitudes are converted to flux densities using the zero magnitudes for Vega, as described in \citet{2006A&A...448..181G}. The IRAS LRS spectrum was taken from the database\footnote{\url{http://isc83.astro.unc.edu/iraslrs/getlrs_test.html}} maintained by Kevin Volk. In general, the errors of the spectroscopic fluxes are scaled typically by a factor of 0.2, to provide roughly equal weight in the photometric and spectroscopic data to the overall data set \citep{2018mnras..479..3545}. The quality of the fit is obtained through a $\chi^2$ analysis. 

\begin{table*}
	\centering
	\caption{Archival photometric data for SED fit}
	\label{tab:archival_phot}
	\resizebox{1.0\textwidth}{!}{
		\begin{tabular}{ccclcllcc}
			
			\hline
			Catalog & $\alpha_{(2000)}$ & $\delta_{(2000)}$ & r*& Filter & Flux (err) & Magnitude (err) & Flag$\dagger$ & Remarks** \\
			& ("h:m:s")         & ("d:m:s")    & (\arcsec)         & (Jy) &  (mag)   &      &  \\
			\hline
			
			2MASS   & 18:30:12.11 & +09:33:42.60 & 0.06 & $J$ & ...  & 7.016 (0.030) & ... & 1 \\
			2MASS   & 18:30:12.11 & +09:33:42.60 & 0.06 & $H$ & ...  & 5.475 (0.027) & ... & 1 \\
			2MASS   & 18:30:12.11 & +09:33:42.60 & 0.06 & $K$ & ...  & 4.517 (0.018) & ... & 1 \\
			AllWISE & 18:30:12.11 & +09:33:42.51 & 0.14 & $W1$ & ... & 3.925 (0.361) & ... & 2 \\
			AllWISE & 18:30:12.11 & +09:33:42.51 & 0.14 & $W2$ & ... & 2.878 (0.431) & ... & 2 \\
			AllWISE & 18:30:12.11 & +09:33:42.51 & 0.14 & $W3$ & ... & 0.503 (0.096) & ... & 2 \\
			AllWISE & 18:30:12.11 & +09:33:42.51 & 0.14 & $W4$ & ... & --0.712 (0.008) & ... & 2 \\
			IRAS    & 18:30:12.15 &	+09:33:41.56 & 1.2  & 12 $\mu$m & 2.37e+01 & 0.595  & 3 & 3 \\
			IRAS    & 18:30:12.15 &	+09:33:41.56 & 1.2  & 25 $\mu$m & 1.84e+01 & --0.719 & 3 & 3 \\
			IRAS    & 18:30:12.15 &	+09:33:41.56 & 1.2  & 60 $\mu$m & 2.24e+00 & --0.386 & 3 & 3 \\
			IRAS    & 18:30:12.15 &	+09:33:41.56 & 1.2  & 100 $\mu$m & 2.24e+00 & ... & 1 & 3 \\
			AKARI   & 18:30:12.10 &	+09:33:43.56 & 0.91 & S09 & 1.067e+01 (4.56e--02) & 1.840 (0.017)   & 3 & 4 \\
			AKARI   & 18:30:12.10 &	+09:33:43.56 & 0.91 & S18 & 1.363e+01 (1.81e+00) & --0.257 (0.136) & 3 & 4 \\
			AKARI   & 18:30:11.88 & +09 33 38.38 & 5.5  & S65 & 5.512e--01  & ... & 1 & 4 \\
			AKARI   & 18:30:11.88 & +09 33 38.38 & 5.5  & S90 & 7.728e--01 (2.52e--01) & --0.268 (0.307) & 3 & 4\\
			AKARI   & 18:30:11.88 & +09 33 38.38 & 5.5  & S140 & ... & ... & ... & 4 \\
			AKARI   & 18:30:11.88 & +09 33 38.38 & 5.5  & S160  & 3.832e--01  & ... & 1 & 4 \\	 
			\hline 
	\end{tabular}} \\
	\textbf{Note.} r* = Distance from center ($\alpha_{(2000)}$ = 18$^h$30$^m$12.10$^s$, $\delta_{(2000)}$ = +09\degr33\arcmin42.6\arcsec) \\ $\dagger$ 1 -- poor quality, 3 -- high quality. Data with Flag = 1 are not considered for the SED fit. \\ ** 1 -- \citet{2003yCat.2246....0}; 2 -- \citet{2013yCat.2328....0}; 3 -- IPAC (1986); 4 -- \citet{2010aap...514A...1} 
\end{table*}

\begin{figure}
\center
    \includegraphics[scale=0.30]{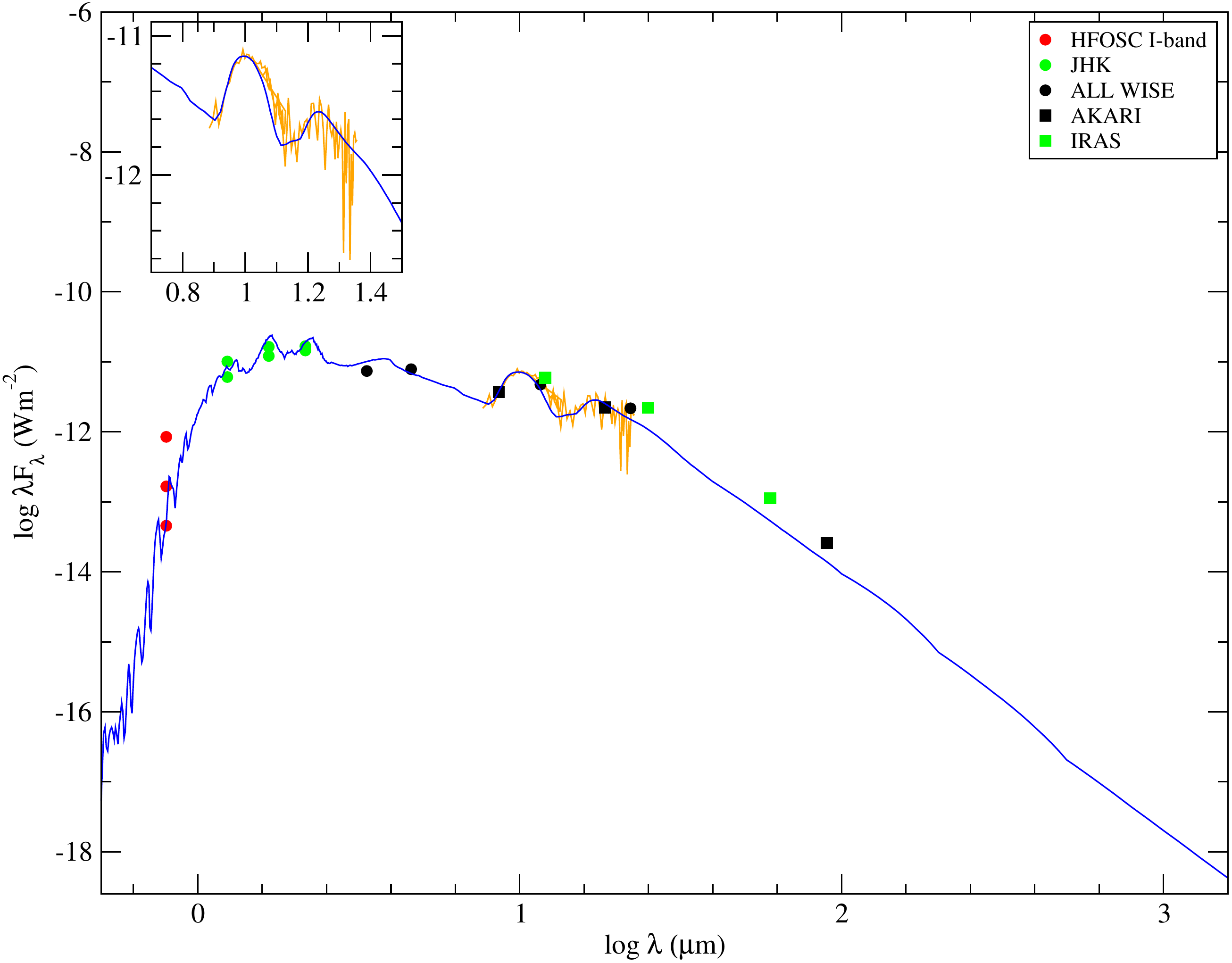}
  \caption{The SED of IRAS 18278+0931 is shown here using multi-wavelength data from optical $I$-band to far-IR, while the inset (top-right) show different data source, e.g., our $JHK$ measurements as well as 2MASS, All WISE, AKARI, IRAS data. The inset in the top-left corner shows the MoD fit of IRAS LRS spectra in zoom.}
 \label{fig2:sed}
\end{figure}  

Input to MoD is a master input file containing the interstellar reddening (A$_V$ = 0.65, \citealt{1998ApJ...500..525}), the distance (D = 4.0 kpc), the path to the file containing the absorption and scattering coefficients of the dust, the path to the file containing the spectrum of the central star, the effective temperature, the number of shells ($N$=1, i.e., one shell model), the outer radius (set to 2000 times the inner radius) and the scaling factors (s(N), set to 1). Finally, initial guesses of the fit parameters ($L$, $\tau_{0.55}$, $T_c$, $p$) are provided, and a code whether these parameters should be varied or fixed (\citealt{2012aap...543..36} for details). The MARCS hydrostatic model atmosphere \citep{2008A&A...486..951} of 2600 K (and log $g$ = 0.0, 1 M$_\odot$, and solar metallicity) is used for the spectra of the central stars. The dust species are taken as mixture of Mg$_{0.8}$Fe$_{1.2}$SiO$_4$:AlO:Fe=100:0:10, calculated using the distribution of hollow spheres with a mean grain size a = 0.2 $\mu$m and a maximum volume fraction of a vacuum core $f_{max}$ = 0.7 $\mu$m. Here, $T_c$ and $p$ have been fixed to 1100 K and 2, respectively, only fitting for $L$ and $\tau_{0.55}$.

 The MoD fit SED is shown in Figure~\ref{fig2:sed}. The $\chi^2$ ($\chi_{red}^2$= 865) of the fit is typically large as the object is a large amplitude variable and non-simultaneous taking of the photometric data from different catalogs. The provided errors are therefore internal errors scaled to a reduced $\chi^2$ of 1. The MoD fit SED provides the resulting parameter, $L$ = 9600 $\pm$ 500$L_{\odot}$ that is comparable to our P-L based estimation, $\tau_{0.55}$ = 9.1 $\pm$ 0.6, inner radius ($R_{in}$) = 5.4 $R_{star}$, MLR ($\dot{M_g}$)= 1.0$\times$10$^{-06}$ $M_\odot$ $yr^{-1}$ and grain density ($\rho$) = 2.65 g cm$^{-3}$. $\dot{M_g}$ is estimated by assuming a 10 km/s expansion velocity and a dust-to-gas ratio of 0.005. The derived MLR a factor 4 lower than the value (3.95$\times$10$^{-06}$ $M_{\odot}$ $yr^{-1}$) that is evaluated by applying the relation
  (${\rm log~(}\dot{M}{\rm )} = -7.37 + 3.42 \times 10^{-3} \times {\rm P}$) 
  provided by \citet{2010A&A...523A..18D} for periods shorter than 850 days. However, considering the large scatter (up to a factor of 10) around the relation in \citet{2010A&A...523A..18D}, we can conclude that the derived values of MLR are in reasonable agreement. Furthermore, the derived MLR appears to be on the low side for an OH/IR star. The low mass-loss rate of the OH/IR star likely implies that the star does not evolve to the superwind phase, a phase the highest mass-loss rates towards the tip of the AGB \citep{1983araa..21..271}.
 
 \begin{figure*}
\center
\includegraphics[scale=0.60, clip=true]{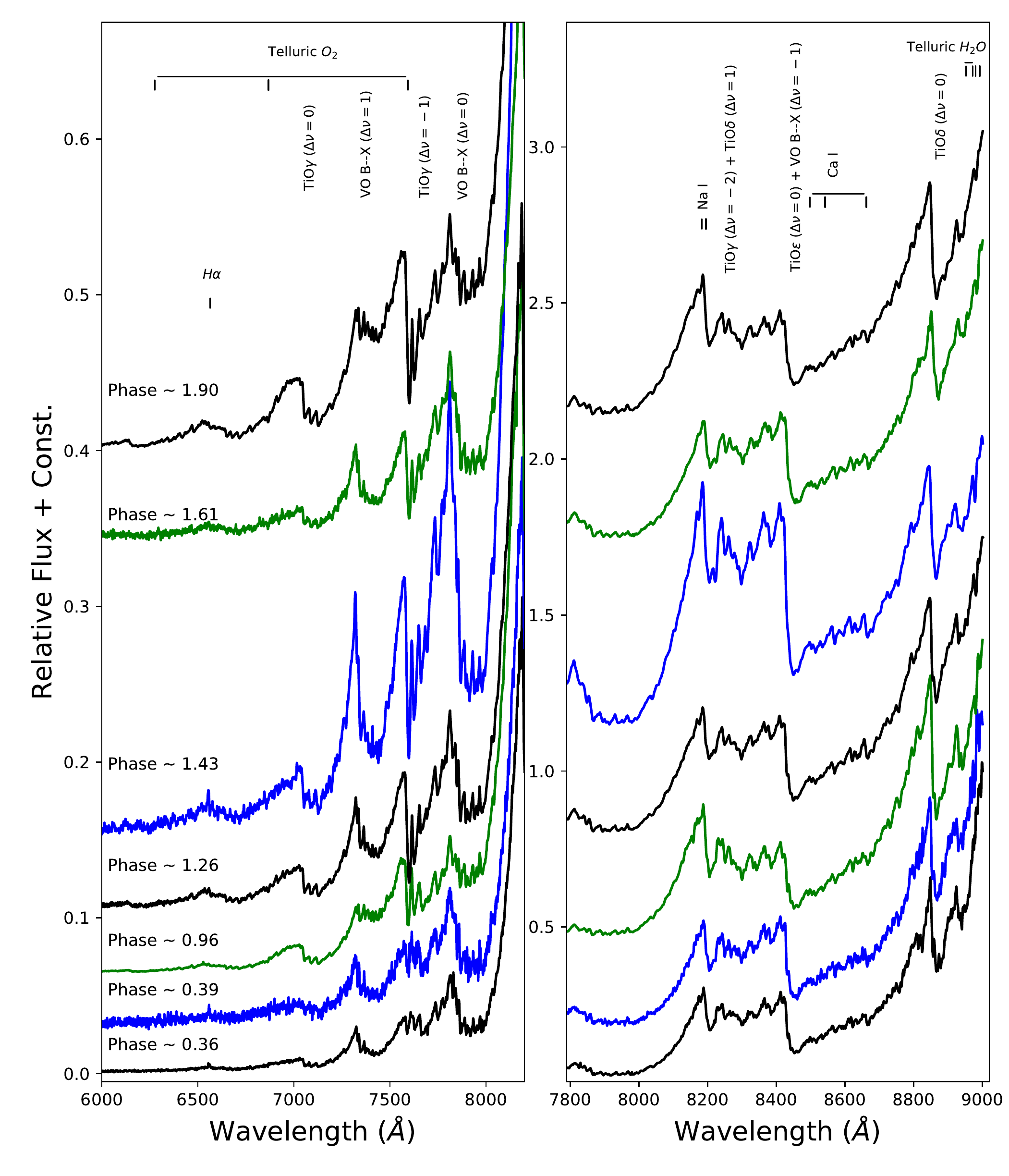}
\caption{The optical spectra of the object IRAS 18278+0931 are presented in the range 6000$-$9000 \AA,~ which show visible features of TiO and VO bands at different phases. The spectra have been normalized to unity at 9165 \AA,~ and offset by constant values 0.0, 0.025, 0.065, 0.105, 0.140, 0.335, 0.40 (left panel) and 0.0, 0.15, 0.42, 0.75, 1.05, 1.70, 2.05 (right panel) respectively with respect to the bottom-most spectra.} 
\label{fig3:optspectra}
\end{figure*}

\begin{figure*}
\center
\includegraphics[scale=0.55, clip=true]{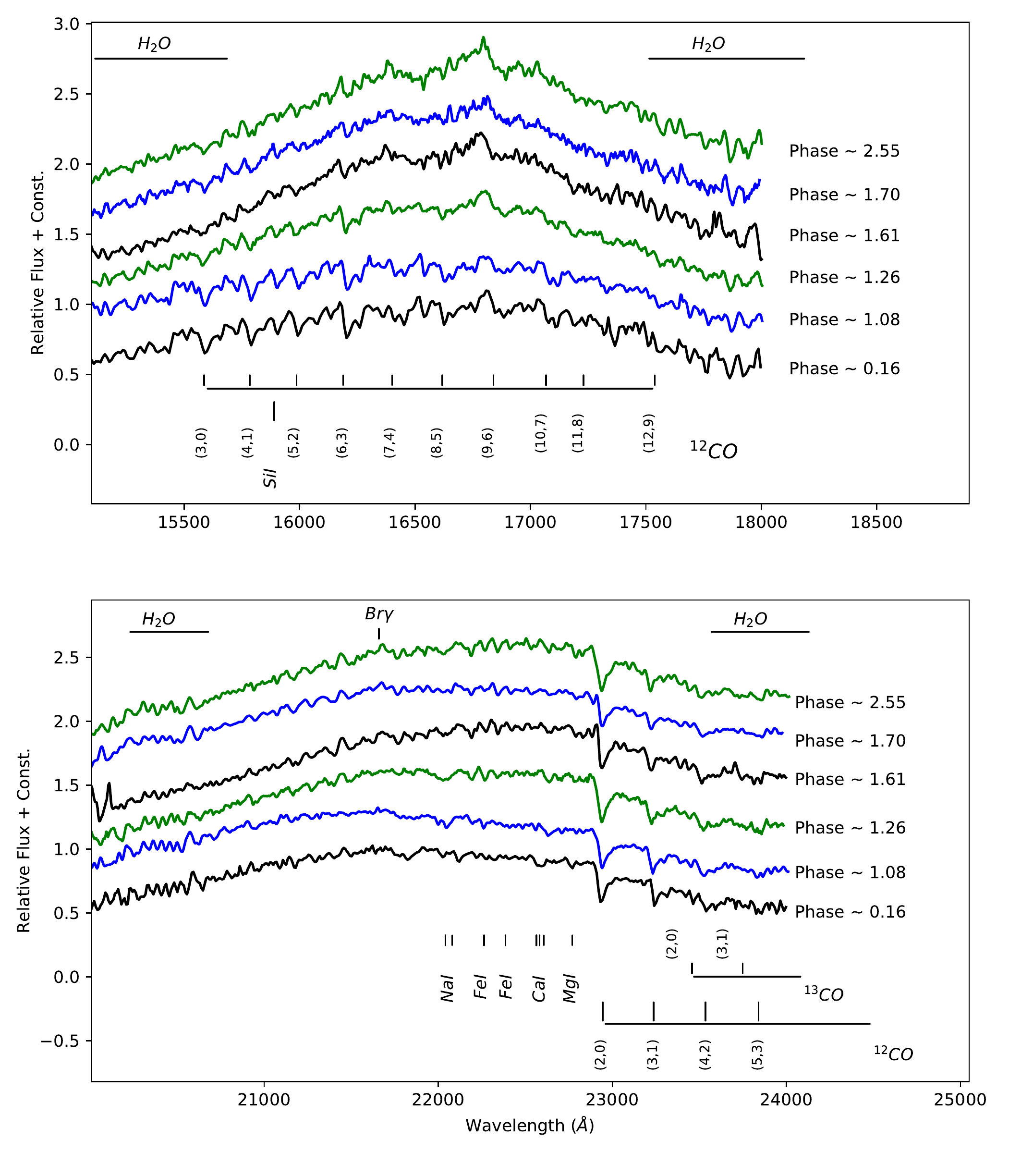}
\caption{The NIR $HK$-band spectra of IRAS 18278+0931 in the wavelength range 1.50$-$2.40 $\mu$m at six different phases are shown in the upper and lower panel, respectively. The NIR spectra at the bottom (phase $\sim$0.16) are taken with the NICMOS-3 instrument on 1.2 m Mt.Abu telescope, and the rest are observed with the TIRSPEC instrument on 2.0-m HCT. The $H$-band spectra in the wavelength range 1.52$-$1.80 $\mu$m show strong four $^{12}$CO second overtone bands including several OH lines. In the $K$-band spectra, the $^{12}$CO first overtone bands are dominated features in the spectra, and Na I, Ca I, and Mg I are seen at 2.20 $\mu$m, 2.26 $\mu$m and 2.28 $\mu$m, respectively. The spectra have been normalized to unity at 16500 \AA~($H$-band), 21700 \AA~($K$-band), and offset by constant value 0.30 with respect to the bottom-most spectra of the same panel.}   
\label{fig4:jhkspectra}
\end{figure*} 

To check how changes in Tc and distance are affecting the output parameters, we change the Tc by 100 K and the distance by 1 kpc. A change in Tc by 100 K changes the derived luminosity by 10 per cent, optical depth by 6 per cent, and MLR by 12 per cent. A change in the distance from 4.0 kpc to 3.0 kpc (5.0 kpc), changes the luminosity and MLR 44 per cent (56 per cent) and 25 per cent (25 per cent), respectively, whereas the optical depth remains unchanged. Thus, we conclude that the derived parameters from the SED fit depend on the choice of input parameters in the model and the accuracy of the distance determination. It is to be noted that the errors for the parameters given for example for the luminosity or optical depth are not reflecting the uncertainty connected to the distance.

Using the distance ($\sim$ 5.0 kpc) provided by the PL relation based on O-rich Mira variables in the LMC (for example, \citealt{2011mnras..412..2345, 2017AJ....154..149Y}), the obtained $L$ from the modelling is a factor $\sim$1.6 larger than the luminosity obtained from the Galactic PL relation of \citet{1991mnras..248..276}. This is because of the overestimation of the distance and thus, the derived distance to the Galactic OH/IR star from the LMC PL relation needs to be considered carefully.

The model estimated luminosity, 9600 $\pm$ $L_\odot$, is consistent with the \citet{1991mnras..248..276} PL relation based estimation as described in Section~\ref{distance_and_luminosity}. We noticed that some of the PL relations (for example, PL relation from \citealt{1990AJ.....99..784, 2011mnras..412..2345}) overestimate the luminosity.

\subsection{Galactic location and current mass} 
Following \citet{1992ApJS...79..105}, the estimated  z-scale height of the object, $\sim$230 pc, suggests the thin disk population \citep{1988aap...200...40, 1992ApJS...79..105, 2008ApJ...673..864}. The bulge population could be ruled out from the IRAS colors criteria (1$<$f$_\nu$(12)$<$5 Jy, 0.5$<$f$_\nu$(12)/f$_\nu$(25)$<$1.5, where $f_\nu$(12) is the flux density at 12 $\mu$m, \citealt{1985A&A...152L...1H}). We thus conclude that our object is an OH/IR star in the Galactic thin disk. Following the mass-luminosity relation in Figure 6 of \citet{1990AJ.....99..784}, the object having $P$ = 575 days and $M_{bol}$ = $-$5.19 lies in the mass ($M$) range of 1$-$1.5 $M_\odot$. Furthermore, the luminosity of our source as estimated in Section~\ref{section:SED}  indicates that it belongs to the ``high-luminosity group". For the ``high-luminosity group", the main-sequence progenitor masses are in the range of 4 M$_\odot$ to 6 M$_\odot$ \citep{2015A&A...579A..76J}. Comparing the current mass of our source to the estimated mass-range of the progenitor, one concludes from the MLR that it has lost mass of 3--5 M$_\odot$ in a mere 3--5 Myr.

\subsection{Optical/NIR spectroscopic studies}
\begin{figure*}
	\center
	\includegraphics[width=15cm,height=15cm,keepaspectratio]{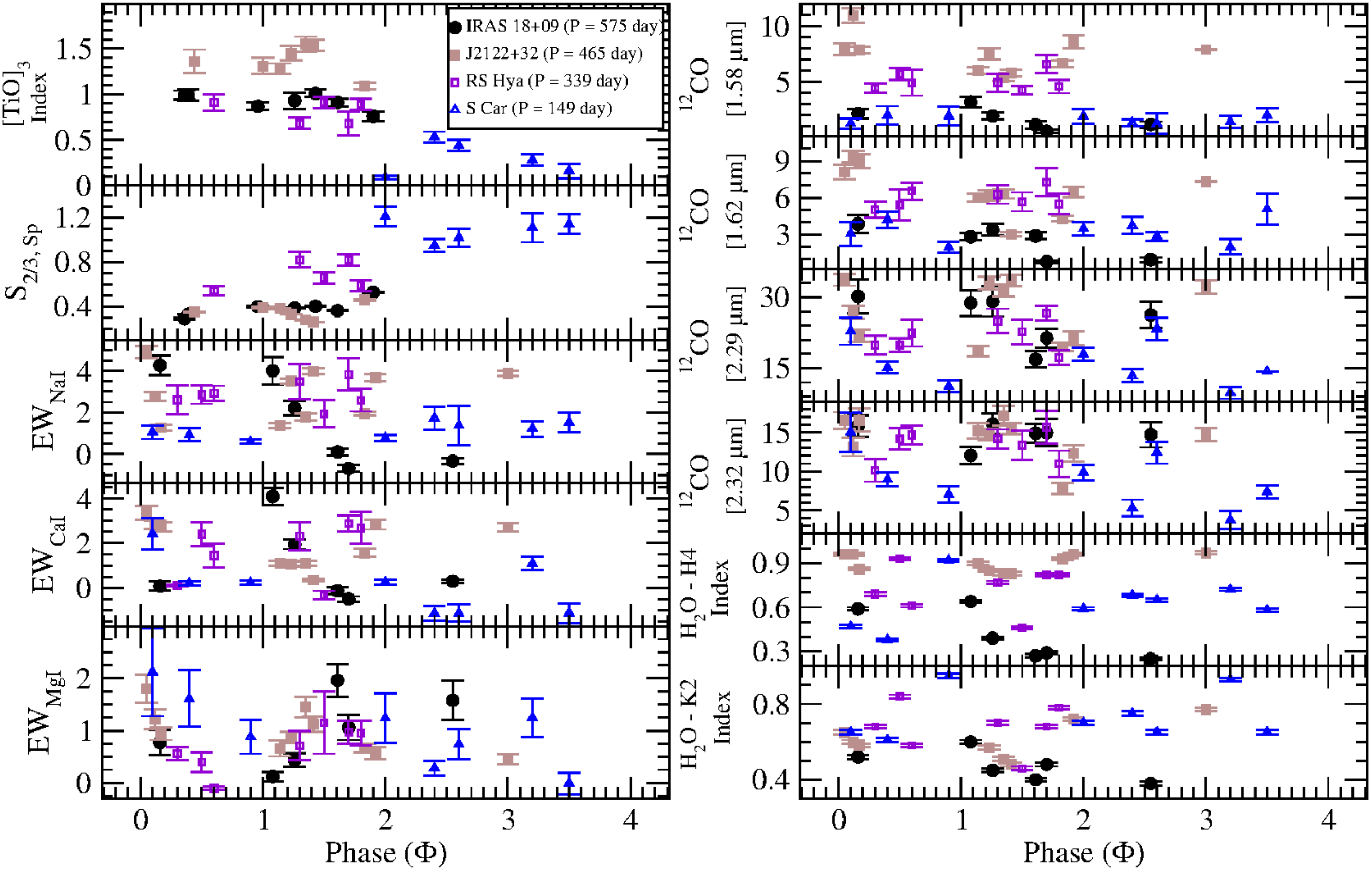} 
	\caption{The phase variation of [TiO]$_3$, S$_{2/3,Sp}$, Na I, Ca I, Mg I, CO, and H$_2$O-K2 equivalent width/index with the visual phase are shown for IRAS 18278+0931, together with MASTER OT J212444.87+321738.3, RS Hya, S Car for comparison. See the text for details.}
	\label{fig5:SiNaCa}
\end{figure*}

 The optical and NIR spectra of the object with variability phases are shown in Figure~\ref{fig3:optspectra} and Figure~\ref{fig4:jhkspectra}, respectively. The spectra show several atomic or/and molecular features as commonly seen in late-type O-rich stars (e.g., \citealt{2000aj..120..2627, 2000aaps..146..217, 2018aj..155..216}). A comparison by-eye shows that all the optical spectra of the object resemble O-rich spectral types later than M6.
 
It is found that the two ends of the $H$ and $K$ spectra bend downwards due to broad H$_2$O absorption features centered at 1.4 $\mu$m, 1.9 $\mu$m \& 2.7 $\mu$m indicating the overall change of the low-resolution continuum shape \citep{2009apjs..185..289, 2018aj..155..216}.

\subsubsection{Phase dependent spectral variability and comparison with Miras} \label{sec:spectral_variability} 

\begin{table}
\caption{Definitions of Spectral Bands }
\label{tab:featureband}
\resizebox{0.5\textwidth}{!}{
\begin{tabular}{cccc}
\hline
\hline
Feature & Bandpass ($\mu$m) & Continuum bandpass ($\mu$m) &   Ref \\
\hline
[TiO]$_3$ & 0.8455$-$0.8725 &  0.8390$-$0.8410, 0.8700$-$0.8725 & 1  \\
Na I	 (2.21 $\mu$m)   & 2.2040$-$2.2107   & 2.1910$-$2.1966, 2.2125$-$2.2170  & 2    \\
Ca I	 (2.26 $\mu$m)   & 2.2577$-$2.2692   & 2.2450$-$2.2560, 2.2700$-$2.2720  & 2   \\
Mg I (2.28 $\mu$m)   & 2.2795$-$2.2845   & 2.2700$-$2.2720, 2.2850$-$2.2874  & 3  \\

$^{12}$CO (1.58 $\mu$m) & 1.5752$-$1.5812 & 1.5705$-$1.5745, 1.5830$-$1.5870 & 4 \\
$^{12}$CO (1.62 $\mu$m) & 1.6175$-$1.6220 & 1.6145$-$1.6175, 1.6255$-$1.6285 & 5 \\
$^{12}$CO (2.29 $\mu$m) & 2.2910$-$2.3020 & 2.24200$-$2.2580, 2.2840$-$2.2910 & 4 \\ 
$^{12}$CO (2.32 $\mu$m) & 2.3218$-$2.3272 & 2.2325$-$2.2345, 2.2695$-$2.2715 & 4\\         
\hline
\hline
\end{tabular}
}
\textbf{Note.} $^1$\citet{1999aaps..140...69}; $^2$\citet{2001AJ..122..1896}; $^3$\citet{2008ApJ...674..194S}; $^4$\citet{2019MNRAS.484.4619G}; $^5$\citet{2008ApJ...674..194S}.
\end{table}

Time-series spectra are obtained to study the variability on the spectra and to understand the dynamical atmosphere because of the stellar pulsation. For that, we explore the pulsational related variations of some of the important spectral features. We estimate different indices and equivalent widths (EWs) of selected features using continuum bands and feature bands as listed in Table~\ref{tab:featureband}. We follow the method as described in \citet{2019MNRAS.484.4619G} using the IDL script\footnote{\url{https://github.com/ernewton/nirew}} \citep{2014AJ....147...20N} for EWs estimation. To compare the spectral behavior between long-period variables having a different period, we over-plot the same indices, estimated from multi-epoch spectra of S Car (P = 149 days), RS Hya (P = 339 days), and MASTER OT J212444.87+321738.3 (P = 465 days) as illustrated in Figure~\ref{fig5:SiNaCa}. The spectra of those Mira variables are taken from \citet{2000aaps..146..217} (S Car and RS Hya) and  \citet{2018aj..155..216} (MASTER OT J212444.87+321738.3). 

\begin{figure}
\center
\includegraphics[width=6cm,height=10cm,keepaspectratio]{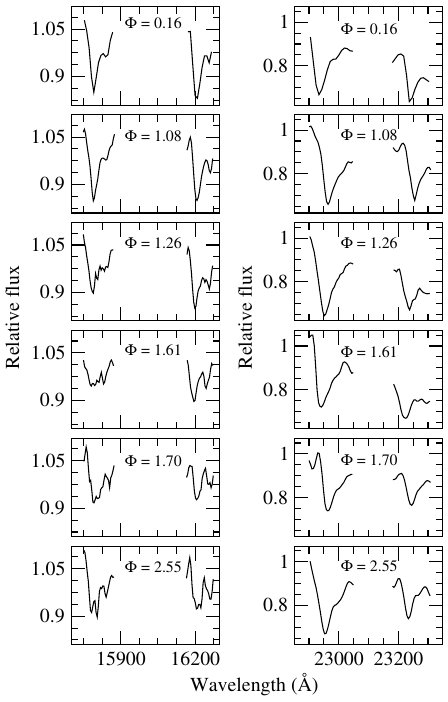} 
\caption{The variable shape of CO-second overtone bandheads (at 1.58 $\mu$m and 1.62 $\mu$m) and CO first overtone (at 2.29 $\mu$m and 2.32 $\mu$m) of IRAS 18278+0931 with phases are shown here.}
\label{fig7:lineshapevariation}
\end{figure}

We explore the optical phase variation of [TiO]$_2$ index centred at 7100 \AA ~\citep{1973AJ.....78.1074}, [TiO]$_3$ index of triple-headed absorption bands at 8433, 8442, 8452 \AA~\citep{1999aaps..140...69} and the flux ratio (S$_{2/3, Sp}$) of two strong absorption bands 770$-$807 nm and 829$-$857 nm (see Figure~\ref{fig3:optspectra}) that is actually related to the spectral type of static giants \citep{1994aaps..105..311}. While the [TiO]$_3$ index shows a small variation with the pulsation phase and increases as the visual brightness decreases, the [TiO]$_2$ index and S$_{2/3, Sp}$ show no significant variation indicating the saturation of TiO bands. However, the VO band (see Figure~\ref{fig3:optspectra}) also contributes towards [TiO]$_3$ index measurement, and a small change in [TiO]$_3$ index with phase can be due to the change in VO band strength.

We have estimated EWs of Na I (2.20 $\mu$m), Ca I (2.26 $\mu$m), Mg I at (2.28 $\mu$m) and CO bands ($^{12}$CO at 1.58, 1.62, 2.29, and 2.32 $\mu$m) as described above, H$_2$O-H4 and H$_2$O-K2 indices\footnote{
 H$_2$O-H4 = $\frac{\langle F (1.531-1.541)\rangle / \langle F (1.670-1.690)\rangle}{\langle F (1.670-1.690)\rangle / \langle F (1.742-1.752)\rangle}$,  \\
H$_2$O-K2 = $\frac{\langle F (2.070-2.090)\rangle / \langle F (2.235-2.255)\rangle}{\langle F (2.235-2.255)\rangle / \langle F (2.360-2.380)\rangle}$;\\
 where $\langle F (a~$--$~b)\rangle$ represents the median flux level in the wavelength range defined by a and b in $\mu$m. H$_2$O-H4 and H$_2$O-K2 indices measure the curvature of the spectra in $H$ and $K$-band, respectively.} following \citet{2018aj..155..216} and \citet{2012apj..748..93}, respectively. All the indices vary significantly over the phase except CO as shown in Figure~\ref{fig5:SiNaCa}. The minimum of metal lines (except Mg I) and CO absorption occurs around phase 0.7 i.e., shortly after minimum light (see Figure~\ref{fig5:SiNaCa}) which may be attributed to the line blurring and veiling \citep{Merrillbook1940, 2000aaps..146..217}. Furthermore, a clear change in the line shape of CO overtone band heads with phase is presented in Figure~\ref{fig7:lineshapevariation}. The reason for line shape variation is the velocity stratification due to the presence of shock \citep{1979ApJ...234..548, 2010aap..514..35}. The appearance of such variability in our low-resolution spectra (Figure~\ref{fig7:lineshapevariation}) are due to modification in the combined shape of several unresolved individual lines as a whole. However, our low-resolution spectra limit the investigation of line doubling and the detail change in velocity profile as discussed earlier \citep{1979ApJ...234..548, 2010aap..514..35}.
 
A significant variation of spectral indices over the phase is evident for individual stars. However, a comparison of these indices among multiple AGB stars yields no unique variation with phase, except possibly Mg I and water indices. Such different behaviour from star-to-star might be because of their different instantaneous atmospheric structure during pulsation, which makes it difficult to explain the cause of different variability for any spectral feature.  Also, a more complete phase coverage of the pulsation cycle is needed to clarify the correlations.
 
We estimate the spectral type (ST) of the OH/IR star quantitatively at different variability phase using the correlation between ST and [TiO]$_3$ index ([TiO] at 8450 \AA) as in \citet{1999aaps..140...69}. The estimated STs at different phases are shown in Table~\ref{tab:parameters}. The ST of the object varies from M7 to M8 over the phase of the pulsation cycle in our limited phase coverage. However, we could not see the large ST variation over the pulsation which is expected for this kind of long-period large amplitude variable stars because of the saturation of [TiO]$_3$ index.

\begin{table*}
\begin{rotatetable*}
\begin{center}
\caption{Phase-dependent variation of spectral features}
\label{tab:parameters}
\resizebox{1.1\textwidth}{!}{
\begin{tabular}{cccccccccccccccc}
\hline
\hline\\
 Date of  & Optical & [TiO]$_3$  & S$_{2/3, Sp}$ & CO & CO & $H_2O$-H4 & Na I & CaI & Mg I & CO & CO & H$_2$O-K2  & Sp. \\
Obs. & Phase & Index  & & 4$-$1 & 6$-$3 & Index &2.20 $\mu$m & 2.26 $\mu$m  & 2.26 $\mu$m & 2$-$0 & 3$-$1 & Index &  $Type^{1}$ \\
\hline
2013 May 28.92 & 0.16 & ...  & ...  & 2.15 $\pm$ 0.41 & 3.86 $\pm$ 0.73 & 0.59 $\pm$ 0.01 & 4.27 $\pm$ 0.47 & 0.08 $\pm$ 0.21 & 0.77 $\pm$ 0.24 & 30.17 $\pm$ 3.61 & 15.85 $\pm$ 2.02 & 0.52 $\pm$ 0.01 & ...\\
2013 Oct 15.52 & 0.36 & 0.99 $\pm$ 0.05 & 0.289 $\pm$ 0.007 & ...  & ...  & ...  & ...  & ...  & ...  & ...   & ...   & ...  & M8 \\
2013 Nov 07.57 & 0.39 & 0.99 $\pm$ 0.06 & 0.330 $\pm$ 0.003 & ...  & ...  & ...  & ...  & ...  & ...  & ...   & ...   & ...  & M8\\
2014 Aug 19.75 & 0.96 & 0.87 $\pm$ 0.04 & 0.398 $\pm$ 0.006 & ...  & ...  & ...  & ...  & ...  & ...  & ...   & ...   & ...  & M7\\
2014 Oct 29.54 & 1.08 & ...  & ...  & 3.18 $\pm$ 0.44 & 2.83 $\pm$ 0.29 & 0.64 $\pm$ 0.01 & 4.01 $\pm$ 0.66 & 4.09 $\pm$ 0.38 & 0.12 $\pm$ 0.09 & 28.80 $\pm$ 2.72 & 12.05 $\pm$ 1.83 & 0.60 $\pm$ 0.01 & ...\\
2015 Mar 02.93 & 1.26 & 0.93 $\pm$ 0.09 & 0.387 $\pm$ 0.003 & 1.95 $\pm$ 0.28 & 3.39 $\pm$ 0.49 & 0.39 $\pm$ 0.01 & 2.21 $\pm$ 0.35 & 1.94 $\pm$ 0.21 & 0.44 $\pm$ 0.13 & 29.12 $\pm$ 3.19 & 16.10 $\pm$ 1.91 & 0.45 $\pm$ 0.01 & M7.5\\
2015 July 05.67 &1.43 & 1.01 $\pm$ 0.04 & 0.401 $\pm$ 0.007 & ...  & ...  & ...  & ...  & ...  & ...  & ...   & ...   & ...  & M8\\ 
2015 Oct 07.69 & 1.61 & 0.91 $\pm$ 0.04 & 0.362 $\pm$0.005 & 1.15 $\pm$ 0.36 & 2.89 $\pm$ 0.28 & 0.27 $\pm$ 0.01 & 0.10 $\pm$ 0.17 & -0.12 $\pm$ 0.15 & 1.96 $\pm$ 0.31 & 16.85 $\pm$ 1.71 & 14.91 $\pm$ 1.23 & 0.40 $\pm$ 0.01 & M7.5\\
2015 Nov 20.55 & 1.70 & ...  & ...  & 0.57 $\pm$ 0.17 & 0.78 $\pm$ 0.12 & 0.29 $\pm$ 0.01 & -0.70 $\pm$ 0.16 &-0.49 $\pm$ 0.12 &1.06 $\pm$ 0.24 & 21.36 $\pm$ 1.97 & 14.97 $\pm$ 1.75 & 0.48 $\pm$ 0.01 & ...\\
2016 Feb 15.98 & 1.90 & 0.76 $\pm$ 0.05 & 0.527 $\pm$ 0.005 & ...  & ...  & ...  & ...  & ...  & ...  & ...   & ...   & ...  & M7\\
2017 Apr 07.92 & 2.55 & ...  & ...  & 1.16 $\pm$ 0.29 & 0.94 $\pm$ 0.19 & 0.25 $\pm$ 0.01 &-0.34 $\pm$ 0.14 & 0.30 $\pm$ 0.08 & 1.58 $\pm$ 0.37 & 26.27 $\pm$ 2.81 & 14.73 $\pm$ 2.14 & 0.38 $\pm$ 0.01 & ...\\	       
\hline
\hline
\end{tabular}
}
\end{center}
 \textbf{Note.} $^{1}$The spectral type has been estimated using the correlation with [TiO]$_3$ Index
  \end{rotatetable*}
\end{table*}

\section{Summary and conclusion} \label{summary_and_conclusion}
We have characterized the time-dependent properties of the OH/IR star, IRAS18278+0931, using long-term optical/NIR photometric and spectroscopic observations. Our main results are summarized as follows.

\begin{enumerate}
\item We have estimated the variability period of 575 $\pm$ 30 days from the best-fit of optical $RI$-band light curves with wavelength-dependent variability amplitudes  $\Delta R \sim$ 4.0 and $\Delta I \sim$ 3.5 mag. From Period-Luminosity (PL) relation, the distance to the source is estimated to be 4.0 $\pm$ 1.3 kpc, which is also consistent with the distance derived from Gaia parallax.

\item Using DUSTY based MoD, we have done the SED fitting the optical to FIR data and LRS IRAS spectral data. The best-fit SED gives (a) a luminosity of 9600 $\pm$ 500 L$_\odot$ for our source, which is in good agreement with the value derived from the PL relation, (b) and optical depth of 9.1 $\pm$ 0.6 at 0.55 $\mu$m, and (c) a MLR of 1.0$\times$10$^{-6}$ M$_\odot$ yr$^{-1}$. The mass loss-rate is also comparable to the estimated value from the period--MLR relation. The current mass of the object lies in the range of 1$-$1.5 $M_\odot$. Furthermore, the estimated luminosity of our source indicates that it belongs to the ``high-luminosity group". Comparing the current mass of the source to the mass-range (4--6 M$_\odot$) of the progenitors for the ``high-luminosity group", one concludes that it has lost mass of 3--5 M$_\odot$ in a mere 3--5 Myr.

\item The spectra of the object at different variable phases are studied covering the optical range from 0.6 to 0.9 $\mu$m and NIR range from 1.5 to 2.4 $\mu$m. Notable variation in spectral features in all atomic and molecular lines (e.g., TiO, Na I, Ca I and Mg I) over phases are seen illustrating the sensitivity of the spectral features to the dynamical atmosphere of the pulsating object. Furthermore, different lines behave differently, signifying their origin in various depths of the very extended atmosphere.

\item A comparative time-series spectral study between the Mira variables and OH/IR star indicates a possible correlation between the index and the phase for the individual stars, however, they show a large dispersion. No unique correlation is seen between spectral indices and the phase, except possibly Mg I and water indices. A complete phase coverage of the pulsation cycle would help to clarify the correlations.
\end{enumerate}


The authors are very much thankful to the anonymous referee for his/her critical and valuable comments, which	helped us to improve the paper. This research work is supported by S. N. Bose National Centre for Basic Sciences under the Department of Science and Technology, Govt. of India. The authors thank the staff of IAO, Hanle and CREST, Hosakote, who made these observations possible. The facilities at IAO and CREST are operated by the Indian Institute of Astrophysics, Bangalore. We acknowledge the usage of the TIFR Near Infrared Spectrometer and Imager (TIRSPEC). We also acknowledge the usage of `MoD' and `Dusty' code. SG is thankful to M. A. T. Groenewegen for helpful discussions and valuable suggestions on DUSTY code and MoD. This research has made use of the SIMBAD database and of the VizieR catalogue access tool, operated at CDS, Strasbourg, France. The original description of the VizieR service was published in Astronomy and Astrophysics Supplement Series (vol. 143, p. 23).  This publication makes use of data products from the Two Micron All Sky Survey, which is a joint project of the University of Massachusetts and the Infrared Processing and Analysis Center/California Institute of Technology, funded by the National Aeronautics and Space Administration and the National Science Foundation. This publication makes use of data products from the Wide-field Infrared Survey Explorer, which is a joint project of the University of California, Los Angeles, and the Jet Propulsion Laboratory/California Institute of Technology, funded by the National Aeronautics and Space Administration.

{\it Software}: IRAF \citep{Tody1986, Tody1993}, TIRSPEC pipe-line \citep{2014JAI.....350006}, More of Dusty (MoD; \citealt{2012aap...543..36}), DUSTY \citep{1997MNRAS.287..799}\\

ORCID iDs: Supriyo Ghosh (\url{https://orcid.org/0000-0003-4640-3369})

\bibliographystyle{aasjournal} 
\bibliography{reference}

\end{document}